\documentstyle[preprint,aps]{revtex}
\begin{document}

\draft

\title{Optimal filters for the detection of\\continuous gravitational waves}

\author{Nadja S. Magalh\~aes}
\address{Instituto Nacional de Pesquisas Espaciais\\
C.P. 515, S\~ao Jos\'e dos Campos - SP, 12201-970, BRAZIL}

\author{Carlos O. Escobar}
\address{Instituto de F\'{\i}sica\\
Universidade de S\~ao Paulo\\
C.P. 20516, S\~ao Paulo - SP, 01452-990, BRAZIL}

\maketitle

\begin{abstract}
We determine the transfer functions of two kinds of filters that can be used in
the detection of continuous gravitational radiation. The first one optimizes
the signal-to-noise ratio, and the second  reproduces the wave with minimum
error. We analyse the behaviour of these filters in connection with actual
detection schemes.
\end{abstract}

\newpage

\section{Introduction}

The detection of gravitational waves (g.w.) is one of the most fascinating and
challenging subjects in Physics research nowadays. Besides checking the General
Relativity theory, the detection of this phenomenon will mark the beginning of
a new phase in the comprehension of astrophysical phenomena by the use of
gravitational wave astronomy.

Although these waves were predicted at the beginning of the
century \cite{Einstein}, the research on their detection only started around
1960,
with the studies of Joseph Weber \cite{Weber}. The major obstacle to this
detection is the tiny amplitude the g.w. have \cite{Thorne}. Even though the
more sensitive detector now operating \cite{Hamilton} is capable to detect
amplitudes near $ h \sim 5 \times 10^{-19}$, this value must be decreased by
several orders of magnitude so that impulsive waves can be detected regularly.

On the other hand, the discovery of pulsars with periods lying in the
milliseconds range stimulated the investigations on the detection of
gravitational waves of periodic origin. Although these waves generally have
amplitudes even smaller than those emitted by impulsive sources, periodic
sources are continuously emitting gravitational waves in space and they can be
detected as soon as the correct sensitivity is reached. Since many of the
resonant mass antennae now operating are designed to detect frequencies near
1000 Hz, the millisecond pulsars will probably be detected if these antennae
ever become sensitive to amplitudes  $ h \leq 10^{-27}$. This value is bigger
if we consider the Crab pulsar ($f \sim 60 Hz$): $ h \sim 10^{-24}$. There is a
resonant mass detector with a torsional type antenna (CRAB IV) being developed
by the Tokyo group \cite{Morimoto} to detect gravitational waves emitted by the
Crab pulsar. This group expects to reach $ h \sim 10^{-22}$ soon.

The main purpose of this paper is a contribution towards the increase in
sensitivity of resonant mass continuous gravitational wave detectors looking at
the use of adequate filters. We study two kinds of filters, the first
optimizes the signal-to-noise ratio (SNR), and is normally used in the
detection of impulsive waves \cite{Price}. The second filter reproduces the
wave
with minimum error. Both filters apparently were not investigated in the
continuous gravitational wave context yet.

\section{The filter that optimizes SNR}

Linear, stationary filters obey the relation

\[
{\cal O}(t) = \int^\infty_{-\infty} k(t') {\cal I} (t-t') dt'.
\]
$k(t)$ is the
impulse response function that characterizes the filter ${\cal K}$ , ${\cal
I}(t)$ is the input at the filter  and ${\cal O}(t)$ is the filter output.

Generally\footnote{We are only considering random, stationary processes in this
work.}
${\cal I}(t)$ has a useful part, $U(t)$, and an unwanted part, $N(t)$:
${\cal I}(t) = U(t) + N(t)$. We have a similar relation for the filter output,
given by ${\cal O}(t) = U'(t) + N'(t)$.

It is well known from noise theory\cite{Zubakov} that the filter ${\cal K}_o$
that optimizes SNR at its
output\footnote{$<f(t)>$ represents the average value of $f(t)$.},

\begin{equation}
SNR \equiv \frac{|U'(t_0)|^2}{<N'^2(t)>},
\label{1}
\end{equation}
must have the following transfer function:

\begin{equation}
K_o(\omega)=  e^{-\imath \omega t_0}
\frac{\tilde{U}^*(\omega)}{S_N(\omega)},
\label{2}
\end{equation}
with

\[
K_o \equiv \int^\infty_{-\infty}  e^{-\imath \omega t} k(t) dt.
\]
$t_0$ is the instant in which the observation takes
place, ${\tilde U}(\omega)$ is the Fourier transform of $U(t)$ (* denotes
complex
conjugation) and $S_N(\omega)$ is the noise power spectrum density:

\[
S_N(\omega) \equiv \int^\infty_{-\infty}  e^{-\imath \omega t} <N(t)
N(t - \tau)> d\tau.
\]

The maximum SNR at the optimal filter output is given by the expression

\begin{equation}
SNR_o = \frac{1}{2\pi} \int^\infty_{-\infty}
\frac{|\tilde{U}(\omega)|^2}{S_N(\omega)}d\omega.
\label{3}
\end{equation}

{}From (\ref{2}) and (\ref{3}) we conclude that a very weak signal will leave
the
filter when the noise is much stronger than the useful signal at the relevant
frequency range.

\section{Quasi-monochromatic signals}

Equation (\ref{2}) is valid as long as  ${\tilde U}(\omega)$ is well behaved.
For example, if $U(t)$  were a strictly monochromatic wave like

\begin{equation}
U(t) = h_0 \cos\omega_0t,
\label{5}
\end{equation}
it would be difficult to build this filter since
${\tilde U}(\omega) = h_0 \delta(\omega - \omega_0)$.

In order to use the optimal filter (\ref{2}) in continuous gravitational wave
detectors we will describe these waves as {\it quasi-monochromatic} useful
signals. It means that the waves that reach the antenna will be of the
form\footnote{We suppose, for simplicity, that the g.w. has only one
polarization, namely ``$+$".}

\begin{equation}
h_{xx}(t) =
\left \{
  \begin{array}{ccccc}
     2 h_0 e^{-at} \cos(\omega_0t+\frac{\pi}{4})  \;\; , t \geq 0 \\
     2 h_0 e^{at} \cos(\omega_0t+\frac{\pi}{4})  \;\; , t \leq 0 \\
  \end{array}
\right.
,
\label{6}
\end{equation}
The constant {\it a} is related to the signal spectral density bandwidth,
$\Delta\omega_h = \frac{\pi a}{2}$, and the corresponding spectral density is
of
the
form\footnote{At this time we admit $\omega$ real and $\omega_0 \geq 0$.}

\[
 S_h(\omega) = \frac{h_0^2}{2} \left [ \frac{2a}{a^2+(\omega-\omega_0)^2} +
\frac{2a}{a^2+(\omega+\omega_0)^2} \right ].
\]
 The signal (\ref{6}) is quasi-monochromatic whenever
$a << \omega_0$, $\omega_0$ being its central frequency. Note that when
$a \rightarrow 0$ we recover (\ref{5}), the monochromatic case.

The continuous gravitational waves emitted by periodic sources
can be regarded as quasi-monochromatic waves. The frequency of the
Crab pulsar, for
example, which is centered near $60 Hz$, has a slow down rate of
$\sim 0.01 Hz/year$. Besides, the orbital motion of the Earth causes a maximum
variation of $\pm 0.03 Hz/year$, and the spinning motion of Earth implies a
maximum variation of $\pm 2 \times 10^{-5} Hz/day$\cite{Tsubonoblair}.

For future use in the optimal filter expression, (\ref{2}), we write the
Fourier transform of the quasi-monochromatic signal (\ref{6}):

\begin{equation}
{\tilde h(\omega)} = \sqrt{2}ah_0 \left [
\frac{1+\imath}{a^2+(\omega-\omega_0)^2} +
\frac{1-\imath}{a^2+(\omega+\omega_0)^2} \right ].
\label{8}
\end{equation}

\section{The mathematical model of the detector}

\begin{figure}[h]
\centering
\begin{picture}(400,250)
\thinlines
\put(0,0){\framebox(400,250){}}
\end{picture}
\caption{Resonant mass detector scheme.}
\label{figure 1}
\end{figure}
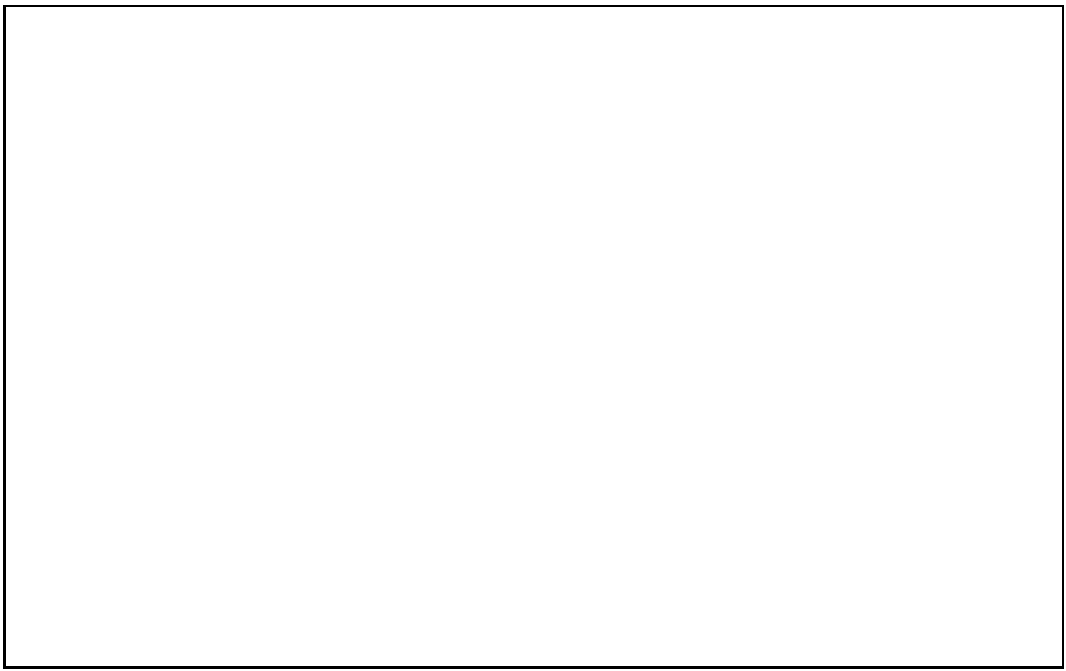

A resonant mass detector can be represented by the scheme of
figure \ref{figure 1}.
In this model $F(t)$ represents the gravitational interaction force between the
g.w. and the antenna. The two-port circuit is related to the massive antenna
and the
transducer\footnote{We will adopt a non-resonant transducer\cite{Pallottino}.},
and it is described by its admittance matrix $y_{ij}(\omega)$, which relates
the force $f_1$ and the velocity  $v_1$  at the input port to the current $I$
and the voltage $V$ at the output port:

\begin{equation}
\tilde{I} (\omega) = y_{11} \tilde{f}_1 (\omega) + y_{12} \tilde{v}_1 (\omega)
\label{9}
\end{equation}

\[
\tilde{V} (\omega) = y_{21} \tilde{f}_1 (\omega) + y_{22} \tilde{v}_1 (\omega)
\]

The transducer and the amplifier have force and velocity noise generators
represented by the stochastic, stationary functions $f(t)$ and $v(t)$,
respectively. $S_f(\omega)$ [$S_v(\omega)$] is the spectral density of
 $f(t)$ [$v(t)$]. We will assume  that these functions are not correlated, so
that  $S_{fv}(\omega) = 0$.

In this model the optimal filter follows the lock-in amplifier. In
figure \ref{figure
2} the elements that precede the optimal filter in the detector are
redrawn\cite{Giffard}.

\begin{figure}[h]
\centering
\begin{picture}(300,200)
\thinlines
\put(0,0){\framebox(300,200){}}
\end{picture}
\caption{Model to the antenna, the transducer and the amplifier.}
\label{figure 2}
\end{figure}
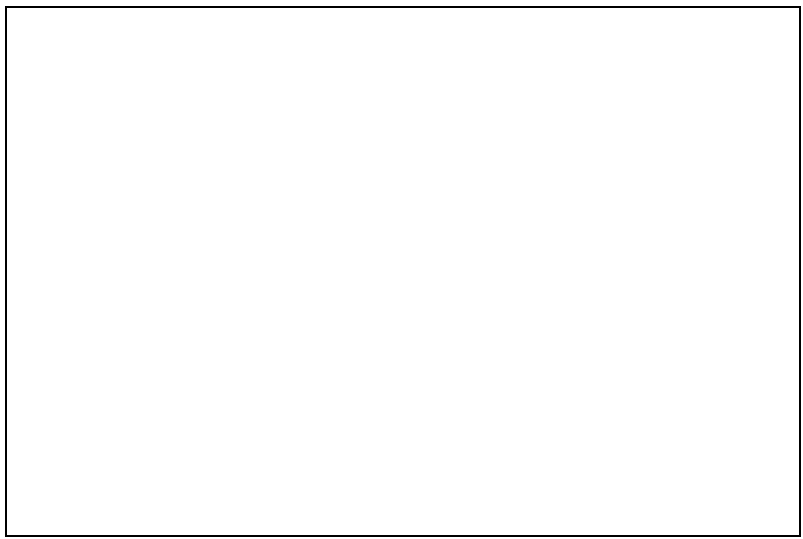

In this figure $m$ is the antenna effective mass, $k$  its elastic constant
and $\mu$
the damping constant. $\phi (t)$  represents the mechanical dissipation at
the antenna, $f(t)$ is associated to the transducer back-reaction on the
antenna and $v(t)$ represents the wideband, serial noise introduced by the
amplifier.

The equation of motion of the system in given by

\[
m \ddot{z}(t) + \mu \dot{z}(t) + kz(t) = - [ \phi (t) + f(t) ] -F(t),
\]
where $z(t)$ represents the displacement. However, in the calculations we will
deal with the velocity ${\dot z}(t)$; at the amplifier output we have
${\dot z}'(t) = {\dot z}(t) + v(t)$.

In the absence of $F(t)$ we can obtain the velocity noise spectral density:

\begin{equation}
S_N(\omega )= \left ( \frac{\tau_0}{2m} \right )^2 \frac{S_\phi (\omega) + S_f
(\omega)}{1+ (\omega \pm \omega_0)^2 \tau^2_0} +
\frac{S_s (\omega)}{1+ (\omega \pm \omega_0)^2 \tau^2_a}.
\label{10}
\end{equation}
At the denominator of this expression, the minus signal is used when
$\omega \geq 0$, and the plus signal is used when $\omega \leq 0$.

Since the thermal and the back-reaction noises are white noises, the force
spectral densities generated by them obey the following Nyquist
relations\cite{MacDonald}:
$S_{\phi}(\omega) = 4 \frac{m k_B T_\phi}{\tau _\phi}$
and
$S_f(\omega) =4 \frac{m k_B T_f}{\tau_f}$.
In these expressions, $\tau_{\phi}$ and  $\tau_f$ are the time constants of
mechanical loss at the antenna and of electrical loss at the transducer and
amplifier, respectively. They are related to the energy decay time of the
antenna, $\tau_0 = \frac{Q_0}{\omega_0}$, according to the expression
$ \frac{1}{\tau_0} =  \frac{1}{\tau_{\phi}} +  \frac{1}{\tau_f}$,
where $Q_0$ is the antenna quality factor. $T_{\phi}$ is the antenna
temperature and $T_f$  is the back-reaction noise temperature. $k_B$ is the
Boltzmann constant.

The function $S_s(\omega)$ that appears in (\ref{10})  represents a serial
white
noise introduced in the useful signal by the electrical network, and it has the
following expression:
$S_S(\omega) = \frac{1}{|y_{22}|^2} 4 R_t k_B T_r$,
where $y_{22}$ comes from (\ref{9}).
$R_t$ is the real part of the impedance at the transducer output and $T_r$ is
the circuit noise temperature.
We assume that the amplifier has a large but limited bandwidth, as it occurs
in practice , given by  $\tau_a^{-1}$. Generally

\begin{equation}
 \tau_0 >> \tau_a.
\label{11}
\end{equation}

Introducing the antenna's {\it equivalent temperature}, $T_e$, defined by the
relation
$\frac{T_e}{\tau_0} = \frac{T_{\phi}}{\tau_{\phi}} + \frac{T_f}{\tau_f}$,
equation (\ref{10}) becomes

\begin{equation}
S_N (\omega) = \frac{\frac{2 \tau_0 k_B T_e}{m} }
{1+ (\omega \pm\omega_0)^2 \tau^2_0} +   \frac{\frac{4 R_t k_B T_r}
{|y_{22}|^2}}{1+ (\omega \pm\omega_0)^2 \tau^2_a}.
\label{12}
\end{equation}
This is the complete expression for the total noise spectral density at the
filter input.

The velocity that the g.w. (\ref{6}) generates at the antenna in the absence of
the noises $\phi(t)$ and $f(t)$ has the following
Fourier transform:

\begin{equation}
{\tilde U}(\omega)=-\frac{\tau_0}{2m}\frac{{\tilde F}(\omega)}{1+\imath
\tau_0 (\omega - \omega_0)}.
\label{13}
\end{equation}
${\tilde F}(\omega)$ is the Fourier transform of the g.w. force on the antenna,
which is given by the relation\cite{Hirakawa?}

\begin{equation}
F(t)=-\frac{1}{4} \sum_{\alpha , \beta =1}^3 q_{\alpha \beta} \frac{d^2}{dt^2}
h_{\alpha \beta}.
\label{13a}
\end{equation}
$q_{\alpha\beta}$ is the dynamic part  of the mass quadrupole tensor of the
antenna. It is a matrix of constant elements which depend on the antenna's
geometry and mass distribution. For our calculations we use
$q_{xx} \sim \rho (\frac{l}{2})^4$,
where $l$ is the characteristic length and $\rho$ is the density of
the antenna.

Equation (\ref{13}) is the useful signal contribution at the filter input,
which corresponds to the spectral
density\footnote{We will simplify the expressions adopting $\omega \geq
0$ henceforth.}

\begin{equation}
S_U(\omega)= \left ( \frac{\tau_0}{2m} \right )^2 \frac{S_F(\omega)}
{1+ \tau_0^2(\omega - \omega_0)^2} .
\label{14}
\end{equation}
$S_F(\omega)$ is the spectral density of the force (\ref{13a}) and it has the
form

\begin{equation}
S_F(\omega)=\frac{1}{4} q_{xx}^2 \omega_0^4 S_h(\omega).
\label{14a}
\end{equation}

\section{The filter that optimizes SNR for the continuous g.w. detector}

Using (\ref{2}), (\ref{12}) and (\ref{13}) and adopting $t_0 = 0$ we obtain the
transfer function of the filter that optimizes  SNR for the model considered
in the preceding section:

\begin{equation}
K_o(\omega)=\frac{\frac{{\cal G} e^{\frac{\imath
\pi}{4}}}{[1+a^{-2}(\omega-\omega_0)^2][1-\imath \tau_0 (\omega -\omega_0)]}}
{\frac{N_p}{1+(\omega-\omega_0)^2\tau_0^2}+\frac{N_s}{1+(\omega-\omega_0)^2
\tau_a^2}}.
\label{18}
\end{equation}
To simplify this expression we introduced the following definitions:
$ {\cal G} \equiv \frac{Q_0 q_{xx}\omega_0h_0}{2ma}$,
$N_p \equiv 2 \tau_0 k_B T_e/m$
and
$N_s \equiv 4 R_t k_B T_r/|y_{22}|^2$.

The maximum SNR related to this filter is obtainable from (\ref{3}), (\ref{12})
and (\ref{14}), and it corresponds to

\begin{equation}
SNR_o=\frac{1}{2\pi} \int_{-\infty}^\infty \frac{\left(\frac{{\cal G}}{1+a^{-2}
(\omega-\omega_0)^2}\right)^2}
{N_p+N_s\frac{1+(\omega-\omega_0)^2\tau_0^2}{1+(\omega-\omega_0)^2\tau_a^2}}
 d\omega.
\label{21}
\end{equation}
Note that when $a << \tau_0^{-1} << \tau_a^{-1}$, this expression becomes

\[
SNR_o \sim \frac{a}{4} \frac{{\cal G}^2}{N_p + N_s}.
\]

Assuming, for simplicity, that
$N_s = N_p$\footnote{We are supposing that the back-reaction
noise and the serial noise give the same contribution to the total signal at
the filter input.},
we find

\begin{equation}
SNR_o \sim \frac{1}{64 k_B} \frac{\omega_0^3 h_0^2}{a} \frac{Q_0 \rho^2 l^8}{m
T_e}.
\end{equation}
By imposing $SNR_o \geq 1$ we obtain the following condition on the parameters
of the detector:

\begin{equation}
\frac{Q_0 \rho^2 l^8}{m T_e} \geq 8.83 \times 10^{-22} \frac{a}{\omega_0^3
h_0^2}.
\label{211}
\end{equation}

To illustrate the use of the filter (\ref{18}) we will adopt two different
realistic detectors. In both cases we will assume that
$N_s = N_p$ and $\tau_0 = 10^2 \tau_a$ (see equation (\ref{11}). The first of
them, detector A, is  designed to detect Crab  pulsar
($\omega_0 = 376.99 Hz$, $h_0 \sim 10^{-24}$ and $a = 3.17 \times 10^{-11} Hz$.
This bandwidth arises from  the slow down of the pulsar after 100 milliseconds
of observation); this detector has the following characteristics:
$Q_0 = 5 \times 10^7$, $m = 1200 kg$\cite{Morimoto}, $T_e = 5 K$,
$\rho=2.74g.cm^{-3}$, $l=1.1m$.

Under these conditions, (\ref{18}) shows a very narrow peak centered in
$\omega_0$. It implies a maximum SNR at its output given by $SNR_o = 1$.
If the signal bandwidth is smaller (implying a smaller observation time),
we have $SNR_o >1$.

On the other hand, if  detector A has a lower equivalent temperature we
attain $SNR_o = 1$ with a longer observation time. For instance, we obtain
this result if
$T_e \sim 0.05 K$, and $a = 3.17 \times 10^{-9} Hz$. Such bandwidth corresponds
to a 10 seconds observation time of the Crab pulsar's slow down .

The other detector considered, detector B, is designed to detect the
millisecond pulsar PSR 1937+214\cite{Pallottino} ($\omega_0 = 8,066.47 Hz$,
$h_0 \sim 10^{-27}$). Since we  did not find any information about the
bandwidth of the g.w. emitted by this pulsar, we will assume that it has the
value $a = 1.2 \times 10^{-8} Hz$. This detector is characterized by
$Q_0 = 5 \times 10^7$, $\rho=2.74g.cm^{-3}$, $l=2m$,
$m = 2300 kg$ and $T_e \sim 0.1 K$; these are typical
values of several ultracryogenic cylindrical antennae.

For detector B, (\ref{18}) also shows a very narrow peak centered in
$\omega_0$, implying $SNR_o = 1$. Like detector A, $SNR_o$ is greater if
the signal bandwidth is smaller.

\section{The filter that reproduces the useful signal with minimum error}

After the g.w. is detected we have to determine its shape with minimum error.
This can be accomplished with the help of an adequate linear filter,
${\cal K}_r$, designed to reproduce the useful signal with the greatest
possible
accuracy (depending on the noise and the useful signal present at its input).
This accuracy is characterized by the mean square error, $<\epsilon^2(t)>$,
which is obtained from the instantaneous reproduction error, $\epsilon(t)$,
defined by

\begin{equation}
\epsilon(t) = {\cal O}(t) - \eta(t).
\label{21a}
\end{equation}

 $\eta(t)$ is the desired signal at the filter output. In a simple filtering
process, as the one we are
considering in this work, $\eta(t)$ must be equal to the useful signal  at the
filter input, $U(t)$.

We obtain the transfer function $K_r(\omega)$ of the filter ${\cal K}_r$ by
imposing
$<\epsilon^2(t)> = <\epsilon^2(t)>_{min}$.  This condition
implies\cite{Zubakov}

\begin{equation}
K_r(\omega) = \frac{S_U(\omega)}{S_U(\omega)+S_N(\omega)},
\label{22}
\end{equation}
supposing there is no cross-correlation between $U(t)$ and $N(t)$. The
corresponding mean square error is

\begin{equation}
\epsilon_r \equiv <\epsilon^2(t)>_{min} = \frac{1}{2\pi} \int^\infty_{-\infty}
\frac{S_U(\omega)S_N(\omega)}{S_U(\omega)+S_N(\omega)}d\omega.
\label{23}
\end{equation}

{}From this equation it is evident that the error becomes smaller if so becomes
the noise. On the other hand, if the noise is too strong ($S_N(\omega)
\rightarrow \infty$)  it results $K_r(\omega) \rightarrow  0$ and no signal
leaves the filter.

If we define the  {\it total} power of a signal $Z(t)$ by
$<Z^2(t)> = \frac{1}{2\pi} \int^{\infty}_{-\infty} S_Z (\omega) d\omega$,
the SNR at the filter input will
be\footnote{Note that (\ref{25}) is different from (\ref{1}). This happens
because we are now interested on the total spectrum of the useful signal, while
in the analysis of the first kind of filter we were interested only on the
maximum amplitude of this signal.}

\begin{equation}
SNR_{\it input} \equiv \frac{<U^2(t)>}{<N^2(t)>}.
\label{25}
\end{equation}

At the filter output the signal  $Z(t)$ will have the following total power:

\begin{equation}
<Z'^2(t)> = \frac{1}{2\pi} \int^\infty_{-\infty} |K_r(\omega)|^2 S_Z(\omega)
d\omega.
\label{24}
\end{equation}
Using this relation we can find the SNR at the filter output,

\begin{equation}
SNR_{\it output} = \frac{<U'^2(t)>}{<N'^2(t)>}.
\label{25a}
\end{equation}

\section{Simple filtering of the quasi-monochromatic signal}

We now consider the particular case of noise spectral density given by
(\ref{12}) and useful signal spectral density given by (\ref{13}). In this case
the filter that reproduces $U(t)$ with minimum error has the following transfer
function:

\begin{equation}
K_r(\omega)=\left(  1+ \frac{\frac{N_p}{1+(\omega-\omega_0)^2\tau_0^2}+
\frac{N_s}{1+(\omega-\omega_0)^2\tau_a^2}}
{\frac{{\cal
G}^2a}{[1+a^{-2}(\omega-\omega_0)^2][1+\tau_0^2(\omega-\omega_0)^2]}}
 \right )^{-1}.
\label{27}
\end{equation}

The minimum error introduced by this filter is obtained from (\ref{23}) using
(\ref{12})  and (\ref{14}):

\begin{equation}
\epsilon _r = \frac{1}{2\pi}\int_{-\infty}^\infty
\left ( \frac{[1+a^{-2}(\omega-\omega_0)^2][1+\tau_0^2(\omega-\omega_0)^2]}
{{\cal G}^2a} +
\frac{1}
{\frac{N_p}{1+(\omega-\omega_0)^2\tau_0^2} +
\frac{N_s}{1+(\omega-\omega_0)^2\tau_a^2}}
\right )^{-1} d\omega.
\label{29}
\end{equation}

In detector A the  total noise power at the filter input
is  $<N^2(t)> = 5.8\times 10^{-24} m^2/s^2$.
If we use  filter  (\ref{27}) in this detector  we obtain $\epsilon_r = 3.42
\times 10^{-31} m^2/s^2$, which is $\sim 6 \times 10^{-8}$ times smaller than
$<N^2(t)>$. Comparing (\ref{25})  and (\ref{25a}) for
this case, we find
$SNR_{\it input} \sim 8.34 \times 10^{-8}$ and $SNR_{\it output} \sim 0.68$, so
that $SNR_{\it output} \sim 8.11 \times 10^6 SNR_{\it input}$.

On the other hand, using filter (\ref{27}) in detector B  we obtain an output
error of $\epsilon_r
=  6.24 \times 10^{-32} m^2/s^2$, which is almost $10^{-6}$ times smaller then
the input noise,  $<N^2(t)> = 6.06 \times 10^{-26} m^2/s^2$.
In this case,
$SNR_{\it input} \sim 1.46 \times 10^{-6}$ and
$SNR_{\it output} \sim 0.67$, which imply $SNR_{\it output} \sim 4.67 \times
10^5 SNR_{\it input}$.

If the Crab bandwidth were $a \sim 1.85\times10^{-11}Hz$ we would obtain
$SNR_{output} \sim 1.05$; this bandwidth also implies $\epsilon_r \sim
2.94\times10^{-32}m^2/s^2$. We would find the same $SNR_{output}$ if the
bandwidth of PSR1937+214 were $a \sim 6.9\times10^{-9}Hz$, corresponding to
$\epsilon_r \sim 5.36\times10^{-32}m^2/s^2$.

\section{Conclusions}

We have derived expressions for the transfer functions of two kinds of filters,
both designed to detect continuous monochromatic waves. The first filter,
${\cal K}_o$, optimizes SNR at its output (equation (\ref{1})) and is important
for a first detection of the wave. The second filter, ${\cal K}_r$, reproduces
the wave with minimum error and should be used when we intend to know the
complete shape of the wave.

In the study of ${\cal K}_o$ we have first analysed the detection of Crab
pulsar. Supposing $N_p = N_s$ (see section $V$) we  concluded that this pulsar
could be
detected if its signal were as monochromatic as
$\sim 3 \times 10^{-11} Hz$; it means a sampling time of the order of
100 msec. We have also
analysed a possible detection of PSR 1937+214 and suggested a maximum limit for
its bandwidth, allowing its detection by  third generation
resonant mass  detectors.

The use of ${\cal K}_r$ is effective in the detection of both pulsars if their
signals are still  more  monochromatic. We then obtain $SNR_{\it output} > 1$
with the form of the signal as preserved as possible.

With the continuous optimization of present detectors the condition of
monochromaticity of the signal becomes weaker. For example, if $N_p > N_s$,
the bandwidths of the signal can be larger than those we obtained in this
paper.
Besides, the inequality (\ref {211}) can be used as a reference to optimize
resonant mass continuous gravitational detectors. Note that continuous sources
with high frequency, small bandwidth and high amplitude are the most favourable
for detection with less improvement of the detector. On the other hand,
the detector should have a small equivalent temperature and
materials with high quality factor and density should be preferred for the
antenna body, which should be as large as possible.

\acknowledgements

N.S.M. thanks CNPq (Bras\'{\i}lia-DF, Brazil) and FAPESP (S\~ao Paulo-SP,
Brazil) for financial support, and
C.O.E. thanks CNPq for partial support. We are grateful to O.D.Aguiar
and C.Frajuca for fruitful discussions.

\end{document}